\documentclass[aps,prl,reprint,showpacs,byrevtex]{revtex4-2}
\let\oldhat\hat
\renewcommand{\hat}[1]{\oldhat{\mathbf{#1}}}
\usepackage[T1]{fontenc}

\usepackage{times,mathptm}
\usepackage[dvips]{graphicx,color}
\usepackage{titlesec}

\begin{document}
\title{Electron-phonon coupling and superconductivity in an alkaline earth hydride CaH$_6$ at high pressures}
\author{Hyunsoo Jeon$^1$, Chongze Wang$^1$, Shuyuan Liu$^1$, Jin Mo Bok$^2$, Yunkyu Bang$^{2,3}$, and Jun-Hyung Cho$^{1,3*}$}
\affiliation{$^1$Department of Physics, Research Institute for Natural Science, and Institute for High Pressure at Hanyang
University, Hanyang University, 222 Wangsimni-ro, Seongdong-Ku, Seoul 04763, Republic of Korea \\
$^2$ Department of Physics, Pohang University of Science and Technology, Pohang 37673, Republic of Korea \\
$^3$ Asia Pacific Center for Theoretical Physics (APCTP),  Pohang-si, Gyeongsangbuk-do 37673, Republic of Korea}
\date{\today}

\begin{abstract}
Recently, an alkaline earth hydride CaH$_6$ having a sodalitelike clathrate structure has been experimentally synthesized at megabar pressures with a maximum $T_c$ of 215 K, comparable to that of a rare earth hydride LaH$_{10}$. Here, based on first-principles calculations, we find that CaH$_6$ exhibits a huge peak in the Eliashberg spectral function ${\alpha}^{2}F$ around the low-frequency region of H-derived phonon modes, in contrast to LaH$_{10}$ having a widely spreading spectrum of ${\alpha}^{2}F$ over the whole frequencies of H-derived phonon modes. It is revealed that the huge peak of ${\alpha}^{2}F$ in CaH$_6$ is associated with an effective electron-phonon coupling (EPC) between low-frequency optical phonons and hybridized H 1$s$ and Ca 3$d$ states near the Fermi energy. As pressure increases, the strengthened H$-$H covalent bonding not only induces a hardening of optical phonon modes but also reduces the electron-phonon matrix elements related to the low-frequency optical modes, thereby leading to a lowering of the EPC constant. It is thus demonstrated that H-derived low-frequency phonon modes play an important role in the pressure-induced variation of $T_c$ in CaH$_6$. Furthermore, unlike the presence of two distinct superconducting gaps in LaH$_{10}$, CaH$_6$ is found to exhibit a single isotropic superconducting gap.

\end{abstract}
\maketitle


\section{I. INTRODUCTION}

Hydride materials under high pressures have recently attracted great attention in condensed matter physics because of their most plausible candidates for the realization of room-temperature superconductivity (SC)~\cite{review-Zurek,review-Eremets}. In such compressed hydrides, H atoms can be "chemically precompressed"~\cite{Hydride_Ashcroft} to become metallic at relatively lower pressures than the solid phase of elemental hydrogen that requires a high pressure over ${\sim}$400 GPa~\cite{MetalicH-Science2017,MetalicH-Nature2020}. The SC in metallic hydrogen has been explained by the Bardeen-Cooper-Schrieffer (BCS) theory, where the superconducting state is represented as a quantum condensate of electron pairs, so-called Cooper pairs, due to electron-phonon interactions~\cite{BCS}. Motivated by the theoretical predictions of high superconducting transition temperature $T_c$ in compressed hydrides~\cite{H3S-Sci.Rep2014,H2S-JCP2014}, experimentalists successfully synthesized sulfur hydride H$_3$S using diamond anvil cell (DAC) techniques~\cite{diamondanvil-Rev2009,diamondanvil-Rev2018} and measured $T_{\rm c}$ around 203 K at ${\sim}$155 GPa~\cite{ExpH3S-Nature2015}. This remarkable experimental realization of a higher $T_c$ record than those of unconventional superconductors such as cuprates~\cite{Rev-cuprate-2006,Rev-cuprate-2015} and pnictides~\cite{Rev-pnictide-2011,Rev-pnictide-2015} triggered a large number of theoretical and experimental studies to search for high-$T_c$ superconductors from compressed hydrides. Most of high-$T_c$ hydrides are rare earth superhydrides XH$_n$ (X = Y~\cite{ExpYH-Oganov,ExpYH-Eremets,ExpYH-Dias}, La~\cite{ExpLaH10-PRL2019,ExpLaH10-Nature2019,Chongze_PRB_2019,Artur_PRB2020}, and Ce~\cite{ExpCeH9-Nat.Commun2019T.Cui,ExpCeH9-Nat.Commun2019-J.F.Lin,ExpCeH-PRL2021,CeH9-hyunsoo,CeH9-PRB}) having H sodalitelike clathrate structures, where each X atom is surrounded by H cage composed of an abnormally large amount of H atoms ($n$ = 6, 9, and 10). Here, the SC of XH$_n$ is driven via strong electron-phonon interaction within the clathrate H networks where the H$-$H bond lengths are close to that of solid metallic hydrogen~\cite{FeH5-Science2017,Shichang-IC}. Among existing superhydrides, LaH$_{10}$ has the highest observed $T_c$ record of 250${\sim}$260 K at a pressure of ${\sim}$170 GPa~\cite{ExpLaH10-PRL2019,ExpLaH10-Nature2019}. Recently, the observation of room-temperature SC was made in carbonaceous sulfur hydride with a $T_{\rm c}$ of 288 K at ${\sim}$267 GPa~\cite{Exp-CSH-Nature2020}. Therefore, the experimental observations of high-temperature SC in surfur-containing hydrides~\cite{Exp-CSH-Nature2020,ExpH3S-Nature2015} and rare earth superhydrides~\cite{ExpYH-Oganov, ExpYH-Eremets, ExpYH-Dias, ExpLaH10-PRL2019, ExpLaH10-Nature2019,Chongze_PRB_2019,Artur_PRB2020, ExpCeH9-Nat.Commun2019T.Cui, ExpCeH9-Nat.Commun2019-J.F.Lin, ExpCeH-PRL2021,CeH9-hyunsoo,CeH9-PRB} have launched a new era of high-$T_{\rm c}$ superconductors.

As another kind of high-$T_c$ hydride having a H clathrate structure, the alkaline earth metal superhydride CaH$_6$ with the high crystalline symmetry of space group $Im\overline{3}m$ [see Fig. 1(a)] was theoretically predicted to exhibit a $T_c$ of 235 K at 150 GPa~\cite{CaH6-PANS2012}. To verify this prediction, Yanming Ma and his coworkers~\cite{CaH6-arXiv2021} have recently synthesized CaH$_6$ by a laser heated DAC technique and measured a maximum $T_c$ of 215 K at 172 GPa. Such an experimental confirmation of the high $T_c$ over 200 K in CaH$_6$ extends the realm to search for room-temperature superconductors in non-rare earth metal superhydrides. To understand how high-temperature SC emerges in CaH$_6$, it is necessary to investigate its salient electronic, bonding, and phononic properties from which large electron-phonon coupling (EPC) is derived. Interestingly, despite similar H clathrate structures, CaH$_6$ shows the drastically different characteristics of the Eliashberg spectral function ${\alpha}^{2}F$ compared to the high-$T_c$ rare earth hydride LaH$_{10}$, as discussed below.

In this article, we study the electronic, bonding, and phononic properties of compressed CaH$_6$ using first-principles density functional theory (DFT) and density functional perturbation theory (DFPT) calculations. It is found that Ca 4$s$ state is nearly unoccupied while Ca 3$d$ states become occupied below the Fermi energy $E_F$ to hybridize with H 1$s$ state, leading to a mixed ionic-covalent bonding between Ca atoms and H cages. This rearrangement of valence electronic states is likely attributed to an increased Coulomb repulsion between Ca atom and its surrounding H atoms at high pressure. Interestingly, in contrast to LaH$_{10}$ having the widely spreading spectrum of ${\alpha}^{2}F$ over the whole frequencies of H phonon modes, CaH$_6$ exhibits a huge peak in ${\alpha}^{2}F$ around the low-frequency region of H-derived phonon modes. This peculiar feature of ${\alpha}^{2}F$ in CaH$_6$ can be explained in terms of a more effective EPC between low-frequency optical phonons and H 1$s$ state hybridized with rather delocalized Ca 3$d$ states. Meanwhile, for LaH$_{10}$, a strong hybridization between H 1$s$ and localized La 4$f$ states~\cite{Liangliang-PRB} near $E_F$ is likely to weaken the enhancement of EPC related to H-derived low-frequency phonon modes. We find that, as pressure increases, the mixed ionic-covalent Ca$-$H bonding is strengthened to decrease the H$-$H bond length, which leads to lowering of the EPC constant via optical phonon hardening as well as the reduction of ${\alpha}^{2}F$. Therefore, H-derived low-frequency phonon modes in compressed CaH$_6$ are of importance for the variation of $T_c$ with respect to pressure. We further find that CaH$_6$ has a single isotropic superconducting gap, in contrast to LaH$_{10}$ having two distinct superconducting gaps~\cite{chongze-LaH10}.

\begin{figure}[htb]
\centering{ \includegraphics[width=8.5cm]{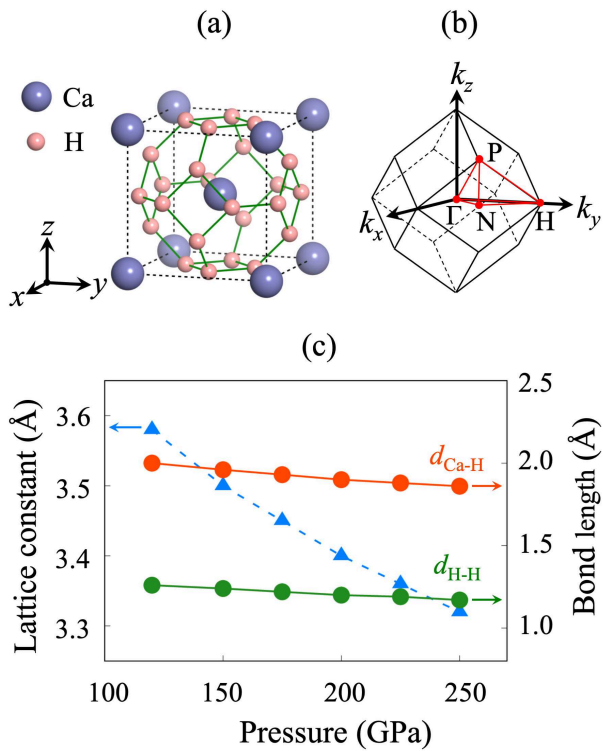} }
\caption{(a) Optimized structure of the $Im\overline{3}m$ phase of compressed CaH$_{6}$, together with the Brillouin zone of the primitive cell. In (b), the calculated lattice constant $a$ = $b$ = $c$ is drawn as a function of pressure, together with the H$-$H and Ca$-$H bond lengths.}
\end{figure}

\section{II. COMPUTATIONAL METHODS}

We performed first-principles DFT calculations using the Vienna {\it ab initio} simulation package~\cite{vasp1,vasp2}. The potential of the core was described by the projector augmented wave method~\cite{paw}. For the exchange-correlation interaction between the valence electrons, we employed the generalized-gradient approximation functional of Perdew-Burke-Ernzerhof~\cite{pbe}. The 3$s^2$3$p^6$ semicore electrons of Ca atom were included in the electronic-structure calculations. A plane-wave basis was used with a kinetic energy cutoff of 400 eV. The Brillouin-zone integration was done with the 24${\times}$24${\times}$24 $k$-point mesh. All atoms were allowed to relax along the calculated forces until all the residual force components were less than 0.001 eV/{\AA}. The lattice dynamics calculation was carried out by using the QUANTUM ESPRESSO package~\cite{QE} with optimized norm-conserving Vanderbilt pseudopotentials~\cite{Dojo} (see Fig. S1 in the Supplemental Material~\cite{supple}) and a kinetic energy cutoff of 1088 eV. Here, we used the $6\times 6\times 6$ $q$-point mesh for the phonon dispersion calculation. For the EPC calculation, we used the software EPW~\cite{epw-iso,epw-aniso,epw-code} with the $24\times 24\times 24$ $q$- and $48\times 48\times 48$ $k$-point meshes in the Brillouin zone.

The essential ingredient of the Eliashberg equations is the Eliashberg function $\alpha^{2}F(\omega)$, defined as:
\begin{equation}
\alpha^2F(\omega) = {1\over 2\pi N(E_F)}\sum_{{\bf q}\nu}
                    \delta(\omega-\omega_{{\bf q}\nu})
                    {\gamma_{{\bf q}\nu}\over\hbar\omega_{{\bf q}\nu}}
\end{equation}
with
\begin{eqnarray}
\gamma_{{\bf q}\nu} &=& 2\pi\omega_{{\bf q}\nu} \sum_{ij}
                \int {d^3k\over \Omega_{BZ}}  |g_{{\bf q}\nu}({\bf k},i,j)|^2
                    \delta(\epsilon_{{\bf q},i} - \epsilon_F) \\\nonumber  &\times& \delta(\epsilon_{{\bf k}+{\bf q},j} - \epsilon_F),
\end{eqnarray}
where $N(E_F)$, $\gamma_{{\bf q}\nu}$, and $g_{{\bf q}\nu}({\bf k},i,j)$ represent the density of states at $E_F$, phonon linewidth, and electron-phonon matrix element, respectively.
The integrated EPC constant $\lambda(\omega)$ is obtained by the integration of ${\alpha^{2} F(\omega)}$:
\begin{equation}
\lambda(\omega)=2\int_{0}^{\omega} d\omega'{\alpha^2 F(\omega')}/{\omega'} ,
\end{equation}
where the total EPC constant is calculated as $\lambda(\omega\rightarrow\infty)$.

\section{III. RESULTS AND DISCUSSION}

We begin by optimizing the structure of compressed CaH$_6$ as a function of pressure using DFT calculations. Based on previous theoretical and experimental studies~\cite{CaH6-PANS2012,CaH6-arXiv2021}, we consider the $Im\overline{3}m$ phase of CaH$_6$ where each Ca atom in the bcc lattice is surrounded by H cage composed of twenty four H atoms [see Fig. 1(a)]. Figure 1(b) shows that the calculated lattice constants of compressed CaH$_6$ decrease monotonously with increasing pressure. Accordingly, the size of H cage is reduced with increasing pressure, therefore decreasing the H$-$H and Ca$-$H bond lengths (denoted as $d_{\rm H-H}$ and $d_{\rm Ca-H}$, respectively), as shown in Fig. 1(b). These changes of $d_{\rm H-H}$ and $d_{\rm Ca-H}$ can be intimately linked to the variation in the H$-$H and Ca$-$H covalent bonding strengths as a function of pressure. Figures 2(a), 2(b), and 2(c) display the charge densities ${\rho}$ of CaH$_6$ at 150, 200, and 250 GPa, respectively. Here, each H-H bond has a saddle point of charge density at its midpoint. The calculated ${\rho}$ values at the midpoints of the H$-$H bond are 0.60, 0.65, and 0.70 $e$/{\AA}$^3$ at 150, 200, and 250 GPa, respectively, indicating that the H$-$H covalent bonding strength increases as pressure increases. Similarly, the covalent character of the Ca$-$H bond is strengthened with increasing pressure, because the ${\rho}$ values at the saddle points [marked ${\times}$ in Figs. 2(a), 2(b), and 2(c)] between Ca and H atoms are 0.33, 0.37, and 0.40 $e$/{\AA}$^3$ at 150, 200, and 250 GPa, respectively. Such pressure-dependent bond lengths and the corresponding covalent bonding strengths in compressed CaH$_6$ affect H-derived phonon frequencies and EPC with respect to pressure, as discussed below.

\begin{figure}[ht]
\includegraphics[width=8.5cm]{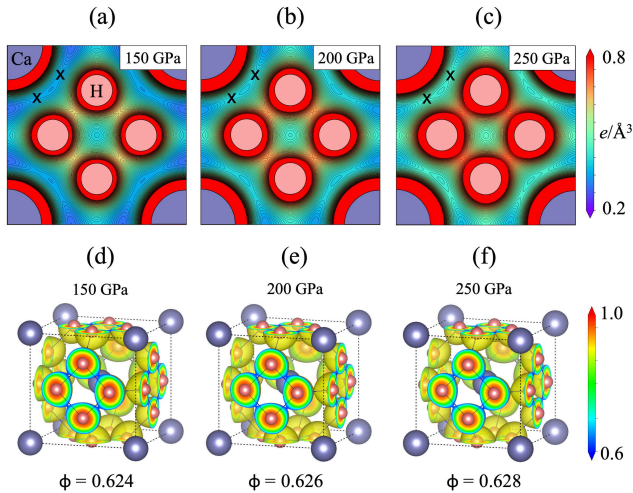}
\caption{Calculated charge densities of CaH$_{6}$ at (a) 150, (b) 200, and (c) 250 GPa. The charge density contour maps are drawn in the (100) plane with a contour spacing of 0.05 $e$/{\AA}$^3$. The charge connections between Ca and H atoms marked "${\times}$" represent the saddle points. The isosurface of electron localization function (ELF) and the corresponding networking value ${\phi}$ are given at (d) 150, (e) 200, and (f) 250 GPa.}
\end{figure}

Figure 3(a) shows the comparison of the electronic band structures of compressed CaH$_6$ at 150, 200, and 250 GPa. We find that the bands below (above) $E_{F}$ shift downward (upward) with increasing pressure, reflecting a band widening due to reduced lattice constants. It is, however, noticeable that the band dispersions around $E_{F}$ change very little with respect to pressure. Consequently, the density of states (DOS) near $E_{F}$ is insensitive to a change in pressure. Such a nearly invariance of electronic states near $E_{F}$ implies that the pressure-induced variations of EPC and $T_c$ are mostly driven by phonon effect, as discussed below. In order to explore the hybridization between Ca and H electronic states, we present the partial DOS (PDOS), obtained at 150 GPa, in Fig. 3(b). We find that the occupied states around $E_F$ are dominantly composed of H 1$s$ and Ca 3$d$ states with a minor component of Ca 4$s$ state. Under high pressure, Ca 3$d$ states that are the excited states in Ca atom become occupied below $E_F$, while Ca 4$s$ state residing further away from the atomic nucleus is pulled above $E_F$ due to an increased Coulomb repulsion between Ca and H atoms. Especially, there are large PDOS peaks of Ca 3$p$ states around $-$10 and $-$22 eV [see Fig. 3(b)], indicating a large delocalization of semicore electrons at high pressure. It is also seen in Fig. 3(b) that (i) the PDOS patterns of Ca 3$p$ and 3$d$ states closely resemble that of H 1$s$ state and (ii) H 1$s$ state is absent in the energy range between $-$17 and $-$21 eV where the gap of Ca-derived states exist. These results indicate a hybridization between H- and Ca-derived states. In Fig. 3(c), we find that the numbers of electrons occupying Ca 3$p$ and 3$d$ states between $-$17 eV and $E_F$ increase with increasing pressure, whereas that occupying Ca 3$s$ state is negligible. This implies that the covalent character of Ca$-$H bond with the hybridization of H 1$s$ and Ca 3$p$/3$d$ states is strengthened with increasing pressure. Meanwhile, our Bader analysis~\cite{Bader} shows that the Bader charges of Ca atom increase as 0.99, 1.07, and 1.14$e$ at 150, 200, and 250 GPa, respectively (see Fig. S2 in the Supplemental Material~\cite{supple}), indicating that the ionic character of Ca$-$H bond due to an electron transfer from Ca to H atoms is weakened with increasing pressure. Based on the results of electronic states and Bader analysis, we can say that the Ca$-$H bond has a mixed ionic-covalent bonding character, similar to the La$-$H bond in LaH$_{10}$~\cite{Seho-PRM}.

\begin{figure}[htb]
\includegraphics[width=8.5cm]{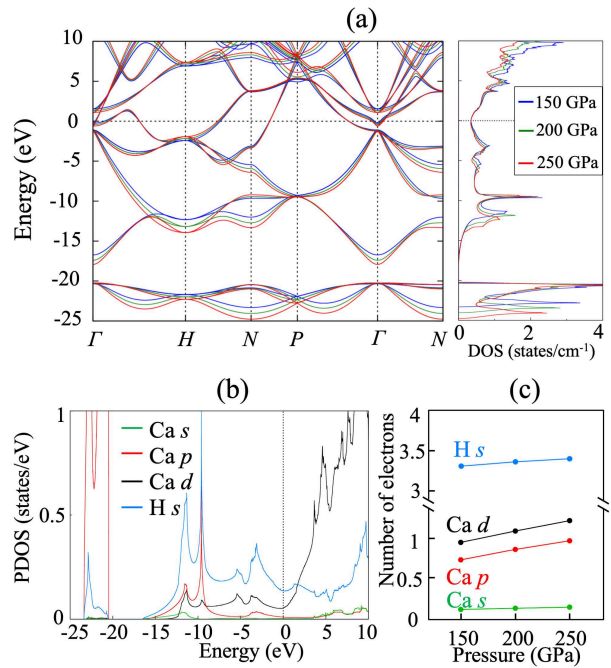}
\caption{(a) Comparison of the electronic band structures of CaH$_{6}$, calculated at 150, 200, and 250 GPa, and (b) PDOS calculated at 150 GPa. The energy zero represents $E_{F}$. In (c), the numbers of electrons occupying H 1$s$, Ca 3$p$, Ca 3$d$, and Ca 4$s$ states between $-$17 eV and $E_{F}$ are plotted as a function of pressure.}
\end{figure}

Next, we examine the phonon spectrum of CaH$_6$ as a function of pressure using DFPT calculations~\cite{QE}. Figures 4(a), 4(b), and 4(c) dispaly the calculated phonon dispersions at 150, 200, and 250 GPa, respectively, together with their projected DOS onto Ca and H atoms. We find that (i) the acoustic phonon modes of Ca atoms are well separated from the optical phonon modes of H atoms and (ii) as pressure increases, H-derived phonon modes shift upwards in the overall frequency range. The logarithmic average ${\omega}_{\rm log}$ of phonon frequencies is estimated to increase as 684, 851, and 933 cm$^{-1}$ at 150, 200, and 250 GPa, respectively. Such pressure-induced phonon hardening can be attributed to the strengthened H$-$H covalent bonding character with increasing pressure. It is noted that at 120 GPa, H-derived low-frequency modes shift towards Ca-derived acoustic modes [see Fig. S3(a) in the Supplemental Material~\cite{supple}]. Such a softening of H-derived low-frequency modes is possibly due to an increase in $d_{\rm H-H}$ as pressure decreases, giving rise to a suppression of restoring forces between H atoms: i.e., increased $d_{\rm H-H}$ gives rise to an effective screening of low-frequency optical modes with the hybridized H 1$s$ and Ca 3$d$ states near $E_F$, as discussed below. At a pressure below 120 GPa, there appear imaginary phonon modes along the $H-N$ line [see Fig. S3(b)], indicating that the $Im\overline{3}m$ phase of CaH$_6$ becomes unstable. It is thus likely that the phononic properties and dynamical instability of compressed CaH$_6$ are intimately correlated with the covalent bonding strength in the Ca-encapsulated H-cage structure via $d_{\rm H-H}$ and $d_{\rm Ca-H}$.

\begin{figure}[htb]
\includegraphics[width=8.5cm]{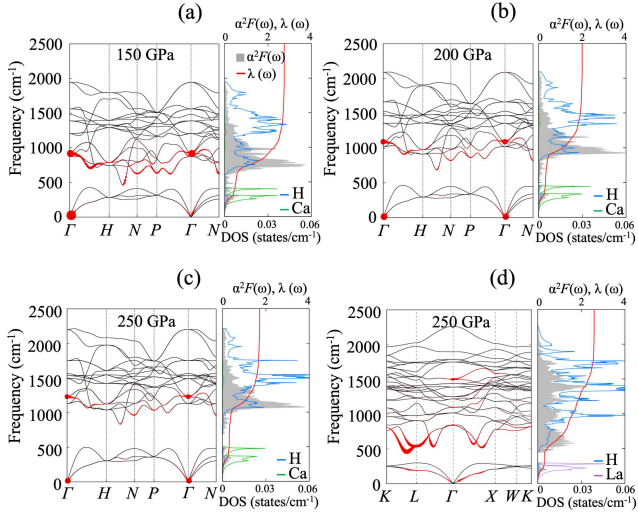}
\caption{Calculated phonon spectrum, phonon DOS projected onto Ca and H atoms, ${\alpha}^{2}F({\omega})$, and ${\lambda}({\omega})$ of CaH$_{6}$ at (a) 150, (b) 200, and (c) 250 GPa. The corresponding results for the $Fm\overline{3}m$ phase of LaH$_{10}$ at 250 GPa are displayed in (d). The size of red circles in phonon spectra is proportional to the EPC strength of each mode.}
\end{figure}

Using the isotropic Migdal-Eliashberg formalism~\cite{Migdal,Eliash,ME-review}, we calculate ${\alpha}^{2}F({\omega})$ and integrated EPC constant ${\lambda}({\omega})$. Figures 4(a), 4(b), and 4(c) include the results of ${\alpha}^{2}F({\omega})$ and ${\lambda}({\omega})$, calculated at 150, 200, and 250 GPa, respectively. We find that ${\alpha}^{2}F({\omega})$ exhibits a huge peak around the low-frequency region of H-derived phonon modes, which in turn yields a steep increase of ${\lambda}({\omega})$. At 150 GPa, the phonon modes between 500 and 1200 cm$^{-1}$ are estimated to contribute to ${\sim}$78\% of the total EPC constant ${\lambda}$ = ${\lambda}$(${\infty}$), while Ca-derived phonon modes reach ${\sim}$14\%. As illustrated by red circles in Figs. 4(a), 4(b), and 4(c), the EPC strength of H-derived low-frequency modes decreases with increasing pressure. We find that ${\lambda}$ values decrease as 2.75, 1.99, and 1.71 at 150, 200, and 250 GPa, respectively. Note that ${\lambda}$ has a maximum value of 4.08 at 120 GPa. Such a lowering of ${\lambda}$ under pressure can be explained in terms of the hardening of optical phonon modes as well as the reduction of the electron-phonon matrix elements [$g_{{\bf q}\nu}$ in Eq. (2)] related to low-frequency optical phonon modes.

It is remarkable that the pattern of ${\alpha}^{2}F({\omega})$ in CaH$_6$ contrasts strikingly with that of LaH$_{10}$ which has an even spread over most frequencies of H phonon modes [see Fig. 4(d)]. As a result, for LaH$_{10}$, ${\lambda}({\omega})$ computed at 250 GPa increases monotonously as ${\omega}$ increases up to ${\sim}$1800 cm$^{-1}$, indicating that most of phonon modes contribute to EPC. These contrasting features of ${\alpha}^{2}F({\omega})$ and ${\lambda}({\omega})$ between CaH$_6$ and LaH$_{10}$ may be attributed to the different aspects of their electronic and bonding properties: i.e., (i) for CaH$_6$, H 1$s$ state hybridizes with rather delocalized Ca 3$d$ states near $E_F$, in comparison with LaH$_{10}$ where H 1$s$ state hybridizes with localized La 4$f$ states~\cite{Liangliang-PRB} and (ii) CaH$_6$ has a longer $d_{\rm H-H}$ than LaH$_{10}$ (see Fig. S4 in the Supplemental Material~\cite{supple}). These aspects of CaH$_6$ can be linked to a more dominant contribution of H-derived low-frequency modes to ${\alpha}^{2}F({\omega})$: i.e., the longer $d_{\rm H-H}$ in CaH$_6$ likely gives rise to larger electron-phonon matrix elements between low-frequency optical phonon modes and hybridized H 1$s$ and Ca 3$d$ states near $E_F$, compared to the case of LaH$_{10}$.

We further study the anisotropy of superconducting gap ${\Delta}$ using the anisotropic Migdal-Eliashberg formalism~\cite{Migdal,Eliash,ME-review}. Figure S5 in the Supplemental Material~\cite{supple} shows the distribution of ${\bf k}$-resolved EPC constant ${\lambda}_{n{\bf k}}$, obtained at 150 GPa. Here, $n$ represents the band index and all available electron-phonon scattering processes connecting the electronic states between ${\bf k}$ and other ${\bf k}$ points are included. With a typical Coulomb pseudopotential parameter of ${\mu}^*$ = 0.13 in superhydrides~\cite{CaH6-PANS2012,rare-earth-hydride-PANS2017,rare-earth-hydride-PRL2017}, we calculate the temperature dependence of ${\Delta}$ at 150 GPa. Figure 5(a) displays the energy distribution of ${\Delta}$ as a function of temperature, indicating that there is a single nodeless superconducting gap. Here, ${\Delta}$ closes at $T_{\rm c}$ ${\approx}$ 240 K, which is almost the same as that (238 K) obtained from the isotropic Migdal-Eliashberg formalism [see the dashed line in Fig. 5(a)]. For $T$ $<$ 100 K, ${\Delta}$ is distributed in the range of 48$-$58 meV, which is correlated with the distribution of ${\lambda}_{n{\bf k}}$ (see Fig. S5 in the Supplemental Material~\cite{supple}). Figure 5(b) shows the $n$- and ${\bf k}$-resolved superconducting gap ${\Delta}_{n{\bf k}}$ on the three Fermi surface sheets arising from the $n$ = 1, 2, and 3 bands, displaying rather uniform superconducting gaps. It is thus demonstrated that CaH$_6$ exhibits a single isotropic superconducting gap, in contrast to LaH$_{10}$ with two distinct superconducting gaps~\cite{chongze-LaH10}.

\begin{figure}[ht]
\includegraphics[width=8.5cm]{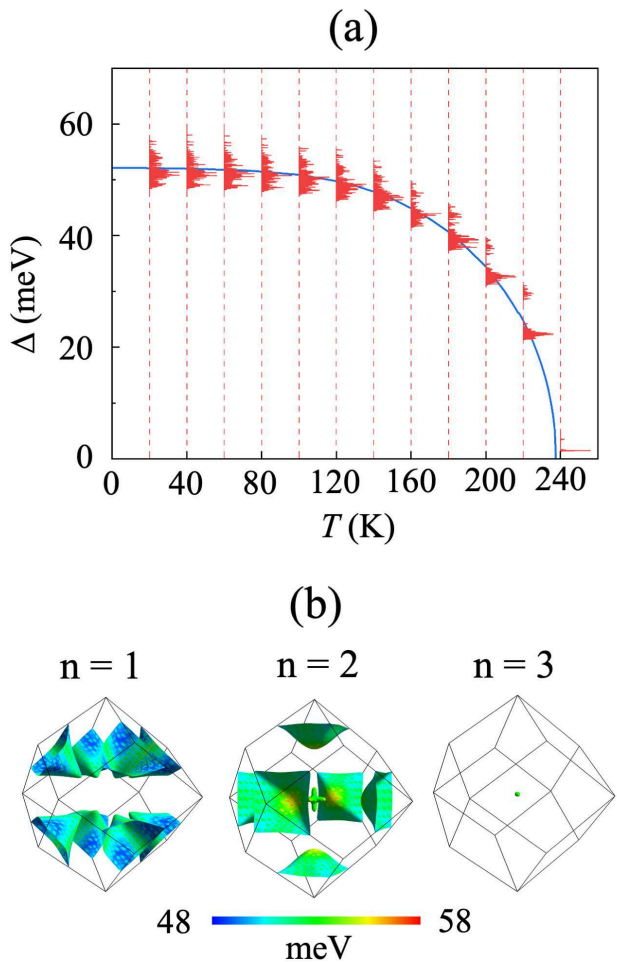}
\caption{(a) Energy distribution of the anisotropic superconducting gap ${\Delta}$ of CaH$_{6}$ at 150 GPa. The solid line in (a) represents ${\Delta}$ values, estimated using the isotropic Migdal-Eliashberg formalism. The \textbf{k}-resolved superconducting gaps ${\Delta}_{n{\bf k}}$ on the three Fermi surface sheets, calculated at 40 K, are drawn in (b). Here, the color scale in
the range between 48 and 58 meV is used.}
\end{figure}

As pressure increases, $T_c$ values estimated using the isotropic Migdal-Eliashberg formalism decrease as 238, 220, and 210 K at 150, 200, and 250 GPa, respectively (see Fig. S6 in the Supplemental Material~\cite{supple}). This lowering of $T_c$ with increasing pressure is determined mainly by a decrease of ${\lambda}$, despite an increase of ${\omega}_{\rm log}$ [see Figs. 4(a), 4(b), and 4(c)]. Recently, Belli $et$ $al$.~\cite{networking} proposed the correlation between electronic bonding network and $T_c$ in superconducting hydrides, where the networking value ${\phi}$ is defined as the highest value of the ELF that creates an isosurface spanning through the whole crystal in all three Cartesian directions. It was claimed~\cite{networking} that the larger the value of ${\phi}$, the higher is $T_c$. In order to examine the ${\phi}$ vs $T_c$ relation, we plot the relevant ELF isosurfaces at 150, 200, and 250 GPa in Figs. 2(d), 2(e), and 2(f), resepctively. Here, ${\phi}$ values are estimated to be 0.624, 0.626, and 0.628 at 150, 200, and 250 GPa, respectively. This increase of ${\phi}$ with increasing pressure is consistent with an increase in ${\rho}$ at the midpoint of H$-$H bond [see Figs. 2(a), 2(b), and 2(c)]. Therefore, we can say that for CaH$_6$, ${\phi}$ increases as pressure increases, but $T_c$ decreases with strengthened H-H bonds at higher pressure.

\section{IV. CONCLUSION}

Using first-principles calculations, we have investigated the EPC and SC in an alkaline earth hydride CaH$_6$ at high pressures. Unlike the high-$T_c$ rare earth hydride LaH$_{10}$ having a widespread of ${\alpha}^{2}F$ over the whole frequencies of H-derived phonon modes, CaH$_6$ exhibits a huge peak in ${\alpha}^{2}F$ around the low-frequency region of H-derived phonon modes. We revealed that such a drastically different feature of the Eliashberg spectral function in CaH$_6$ is associated with an effective EPC between low-frequency optical phonons and H 1$s$ state hybridized with Ca 3$d$ states. Furthermore, as pressure increases, the covalent H$-$H bond is strengthened to yield optical phonon hardening as well as the reduction of ${\alpha}^{2}F$, resulting in a lowering of ${\lambda}$. It was thus demonstrated that H-derived low-frequency phonon modes play an important role in the pressure-dependent variation of $T_c$ in compressed CaH$_6$. The present findings have important implications for understanding the EPC and SC in alkaline earth superhydrides, which can shed light on the discovery of room-temperature superconductors in non-rare earth metal superhydrides.

\vspace{0.4cm}

\noindent {\bf Acknowledgements.}
This work was supported by the National Research Foundation of Korea (NRF) grant funded by the Korean Government (Grants No. 2019R1A2C1002975, No. 2016K1A4A3914691, and No. 2015M3D1A1070609). The calculations were performed by the KISTI Supercomputing Center through the Strategic Support Program (Program No. KSC-2020-CRE-0163) for the supercomputing application research.  \\

H. J. and C. W. contributed equally to this work.


\noindent $^{*}$ Corresponding author: chojh@hanyang.ac.kr

\newpage

\onecolumngrid
\newpage
\titleformat*{\section}{\LARGE\bfseries}

\renewcommand{\thefigure}{S\arabic{figure}}
\setcounter{figure}{0}

\vspace{1.2cm}

\section{Supplemental Material for “Electron-phonon coupling and superconductivity in an alkaline earth hydride CaH$_6$ at high pressures”}
\vspace{1.2cm}
\begin{flushleft}

{\bf 1. Calculated band structures of CaH$_6$ using two different pseudopotentials.}
\begin{figure}[ht]
\includegraphics[width=7cm]{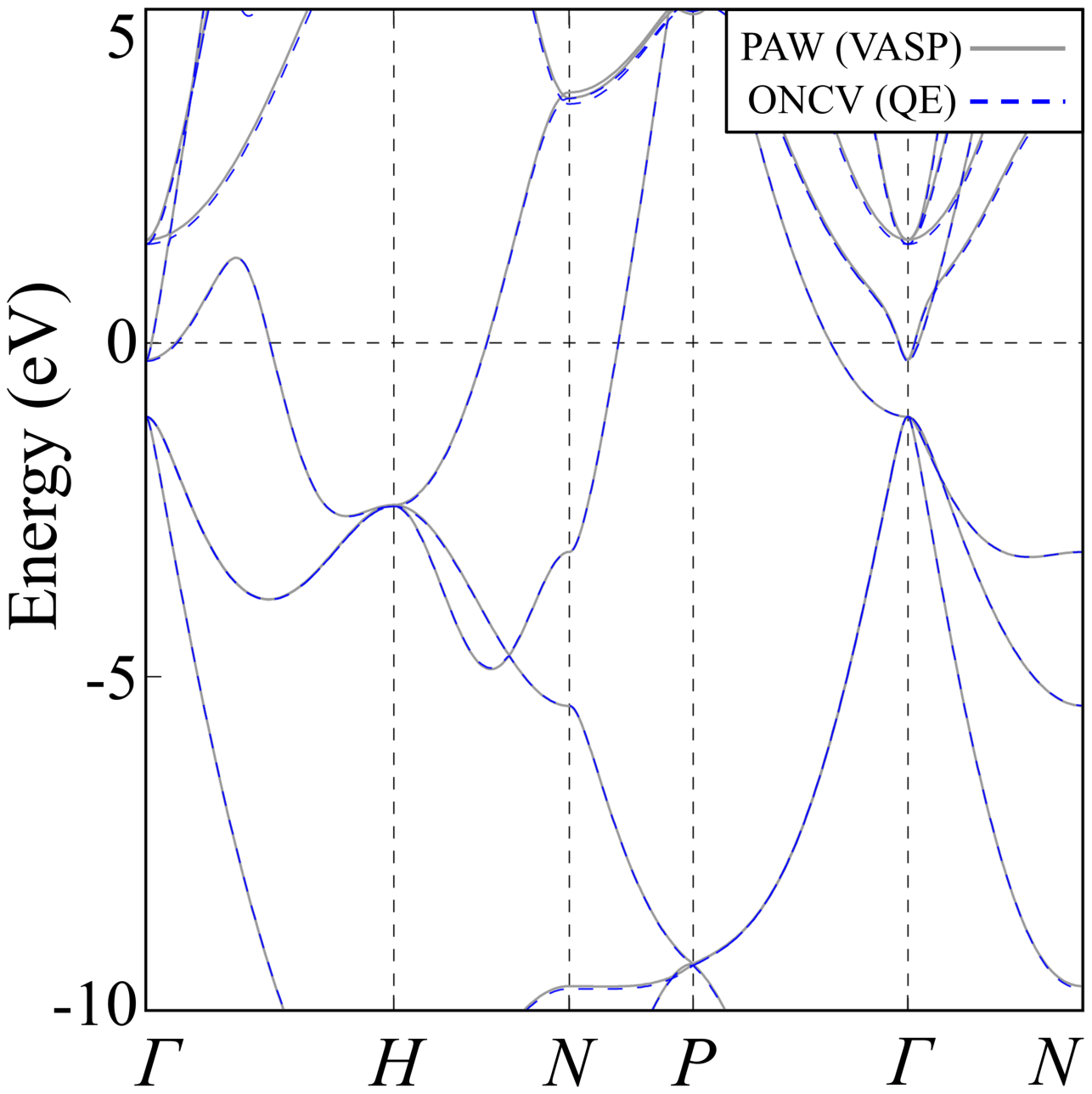}
\caption{ Comparison of the band structures of CaH$_6$, calculated at 150 GPa using PAW pseudopotentials in the Vienna $ab initio$ simulation package (VASP) code and optimized norm-conserving Vanderbilt (ONCV) pseudopotentials in the QUANTUM ESPRESSO (QE) code. We find that (i) the lattice constants optimized using two different pseudopotentials are deviated by less than 0.012\%  and (ii) the two band structures are nearly the same with each other. }
\end{figure}

\vspace{1.2cm}

{\bf 2. Bader charge of Ca atom in CaH$_6$ as a function of pressure.}
\begin{figure}[ht]
\includegraphics[width=13cm]{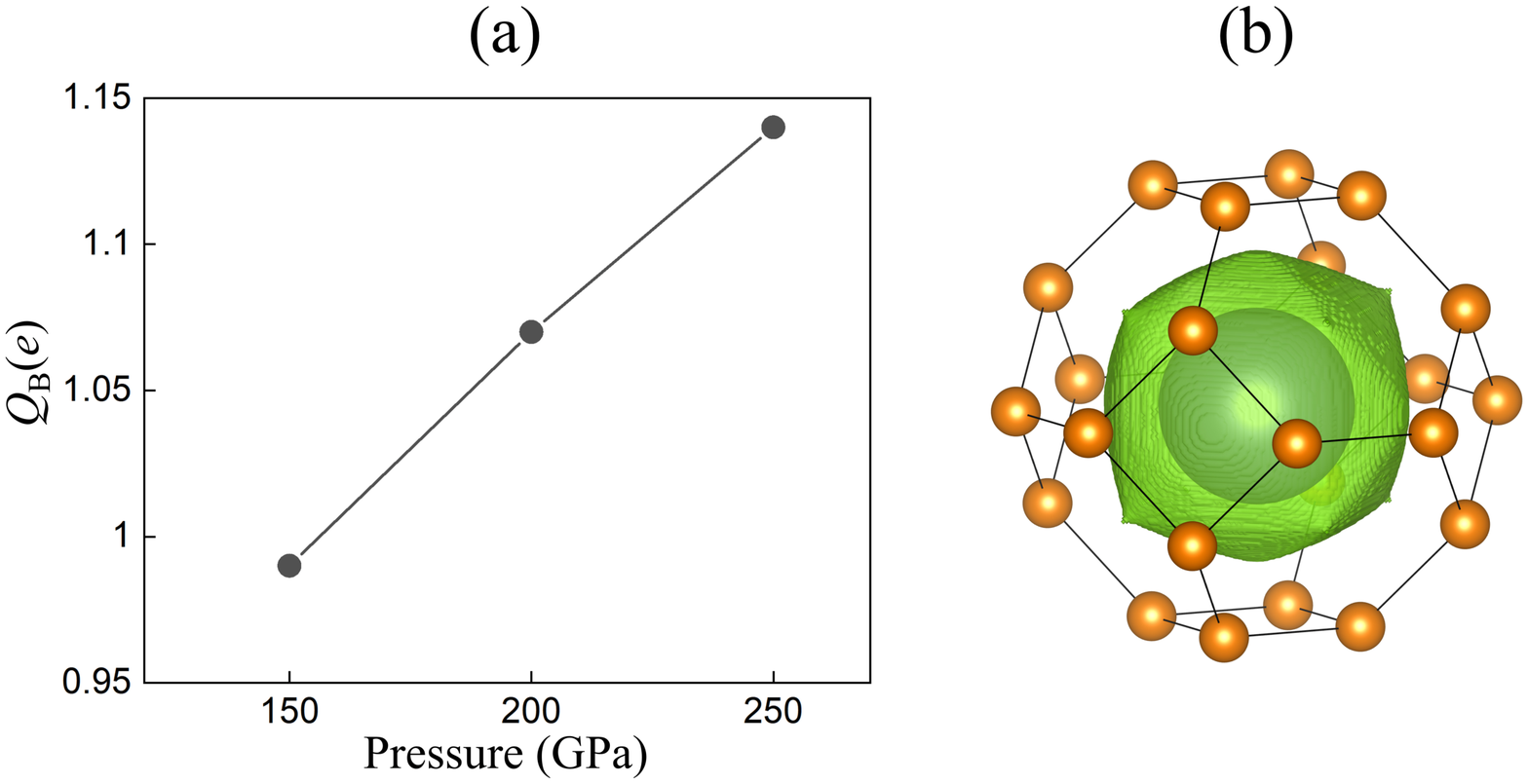}
\caption{(a) Bader charge ($Q_{\rm B}$) of Ca atom in CaH$_6$ as a function of pressure and (b) the Bader basin of Ca atom at 150 GPa.}
\end{figure}
\vspace{1.2cm}

\newpage

{\bf 3. Calculated phonon spectra of CaH$_6$}
\begin{figure}[ht]
\includegraphics[width=13cm]{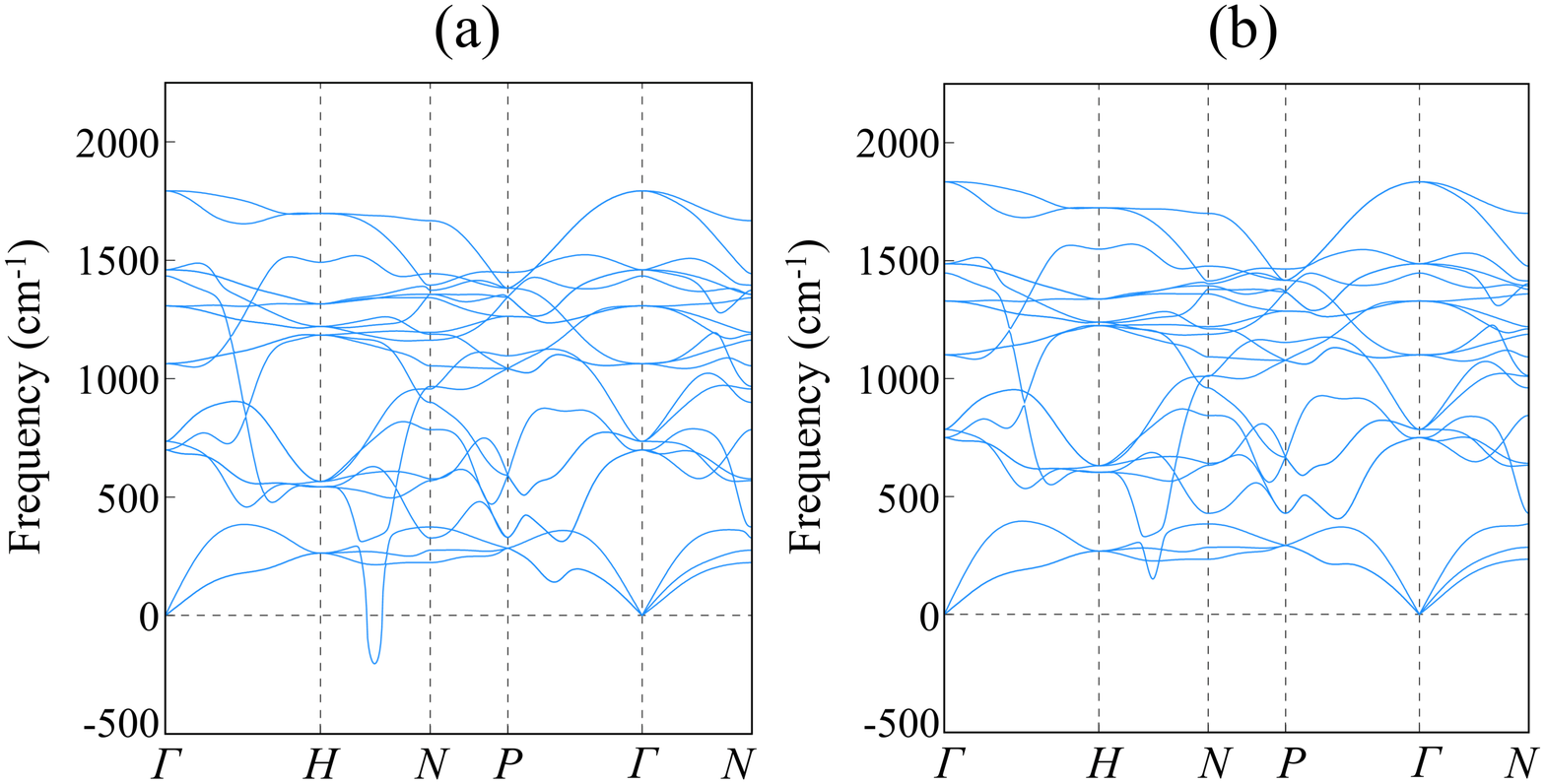}
\caption{ Calculated phonon spectra of CaH$_6$, obtained at (a) 110 GPa and (b) 120 GPa. The structure of CaH$_6$ is thermodynamically stable at 120 GPa without the existence of imaginary phonon frequencies. However, the structure of CaH$_6$ becomes unstable at 110 GPa with imaginary phonon modes.}
\end{figure}

\vspace{1.2cm}

{\bf 4. Comparison of the H-H bond lengths between CaH$_6$ and LaH$_{10}$.}
\begin{figure}[ht]
\includegraphics[width=8cm]{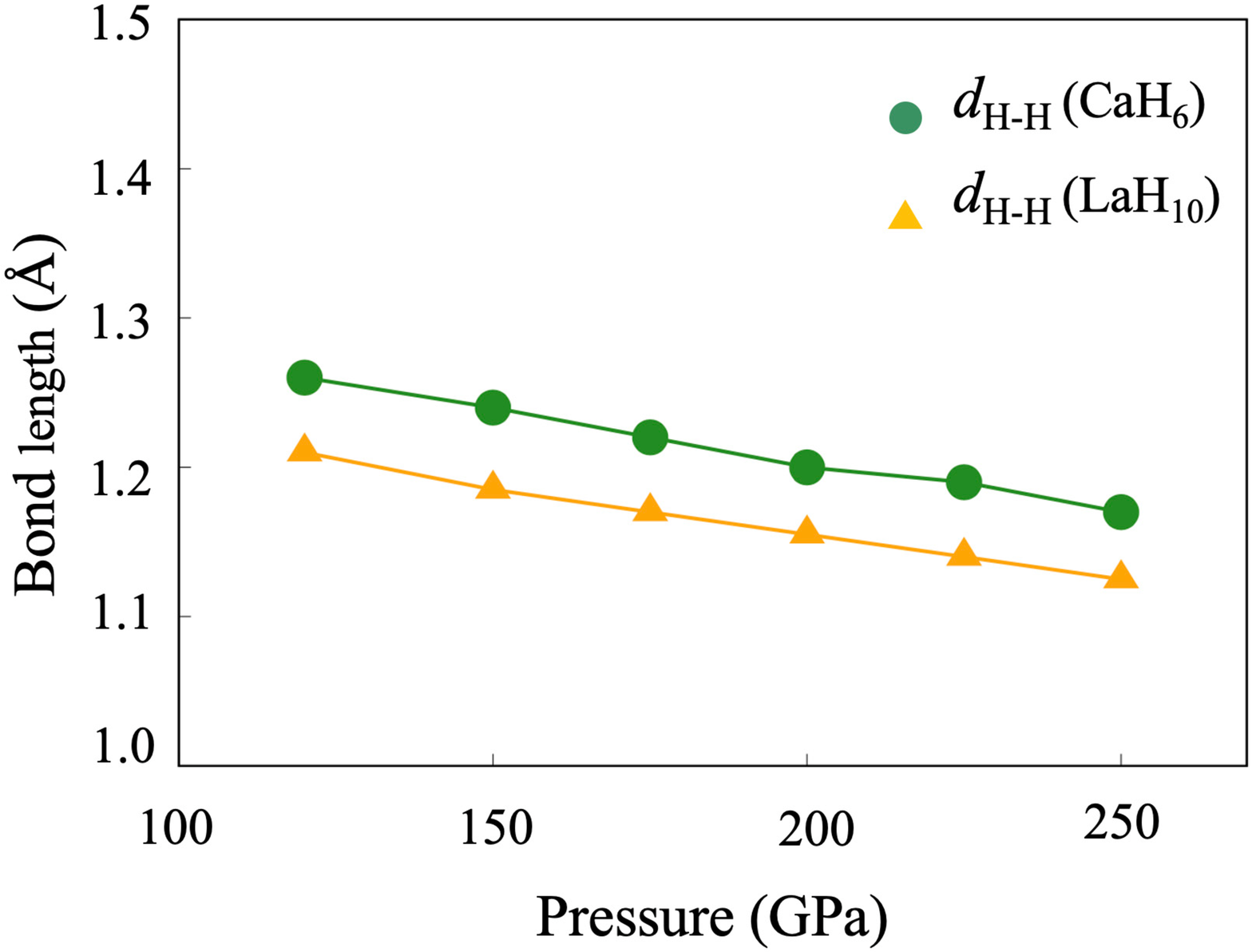}
\caption{ Calculated H$-$H bond lengths ($d$$_{\rm H-H}$) in CaH$_6$ and LaH$_{10}$ as a function of pressure. For LaH$_{10}$, the averaged values of two different H$-$H bond lengths are plotted.}
\end{figure}

\vspace{1.2cm}
\newpage

{\bf 5. Distribution of k-resolved EPC constant ${\lambda}_{n{\bf k}}$.}
\begin{figure}[ht]
\includegraphics[width=7cm]{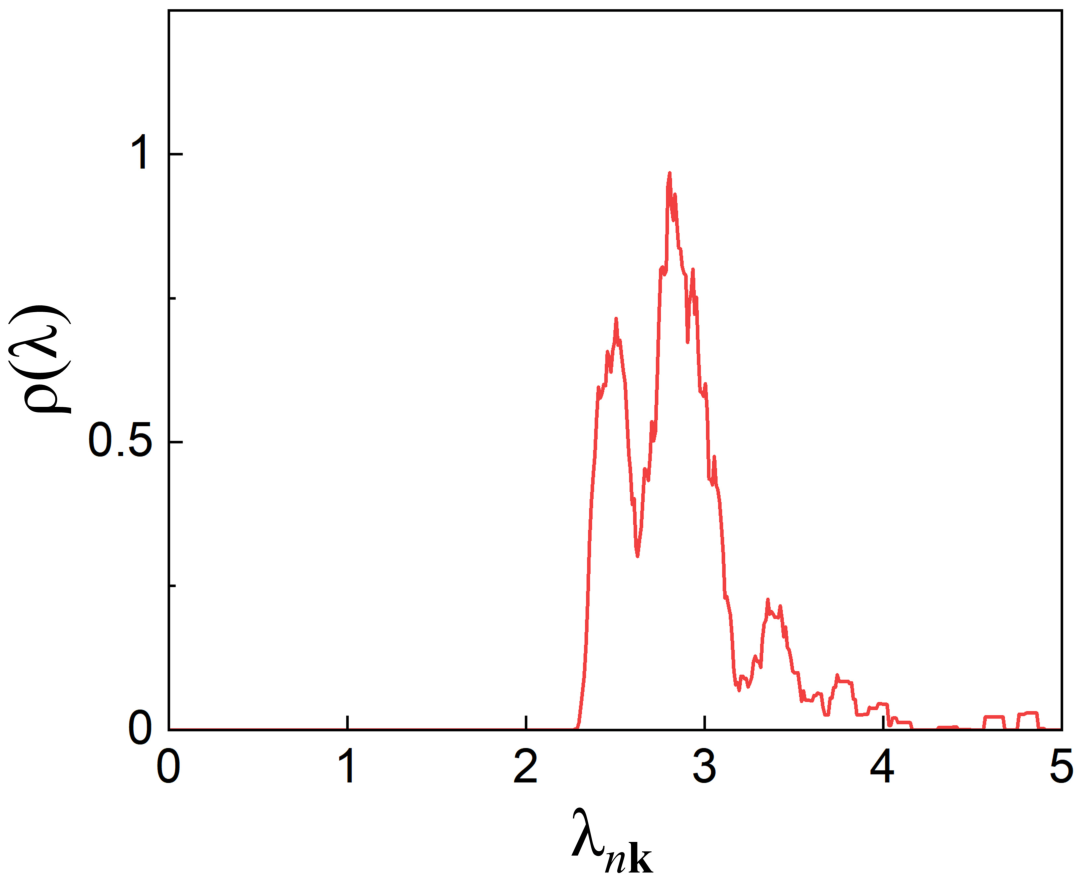}
\caption{ Distribution of {\bf k}-resolved EPC constant ${\lambda}_{n{\bf k}}$, obtained at 150 GPa. }
\end{figure}

\vspace{1.2cm}
{\bf 6. Isotropic superconducting gap of CaH$_6$ at different pressures.}
\begin{figure}[ht]
\includegraphics[width=9cm]{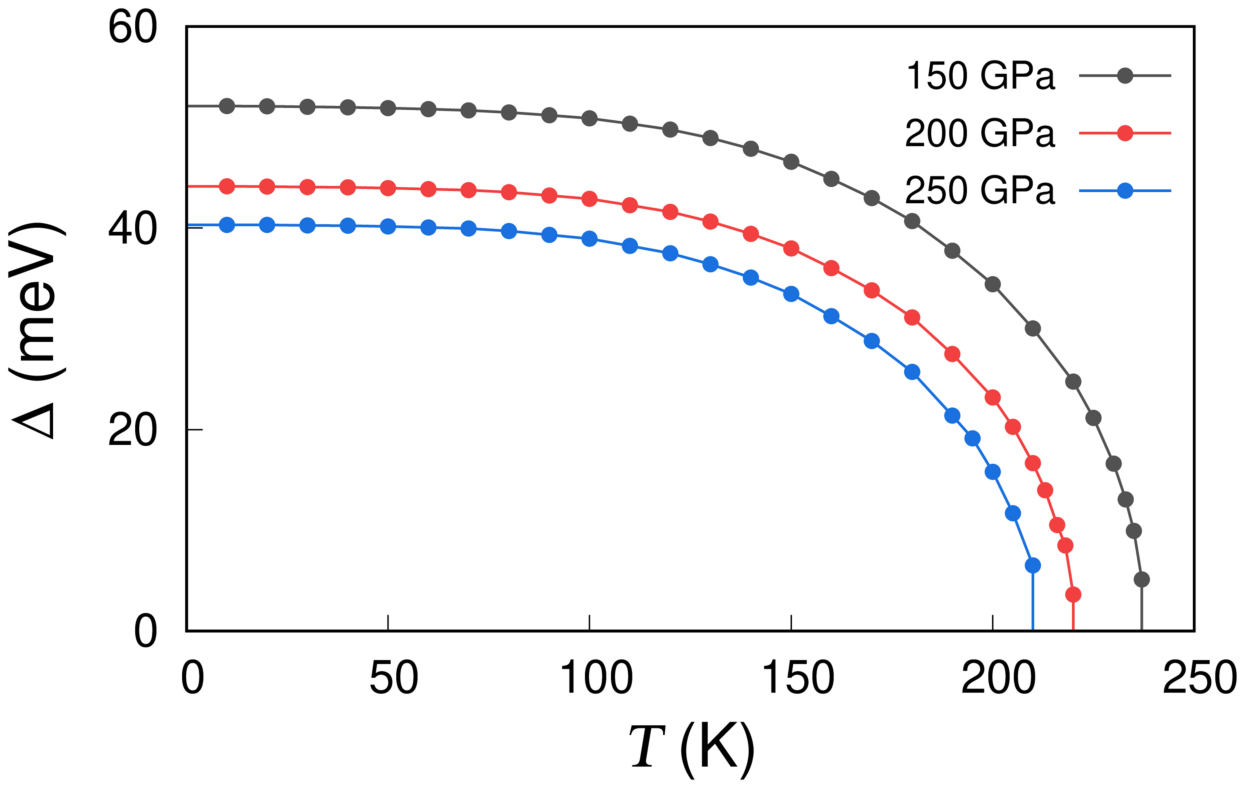}
\caption{ Isotropic superconducting gap ${\Delta}$ of CaH$_6$ as a function of temperature $T$, obtained at 150, 200, and 250 GPa. Here, we used the typical Coulomb pseudopotential parameter of ${\mu}^*$ = 0.13. }
\end{figure}

\vspace{1.2cm}

\end{flushleft}
\end{document}